\documentclass[paper=a4, fontsize=11pt]{scrartcl}
\usepackage[T1]{fontenc}
\usepackage{fourier}

\usepackage[english]{babel}								
\usepackage[protrusion=true,expansion=true]{microtype}	
\usepackage{amsmath,amsfonts,amsthm}
\usepackage[pdftex]{graphicx}	
\usepackage{url}
\usepackage{siunitx}

\usepackage[pdfencoding=auto]{hyperref}
\usepackage{subcaption}
\usepackage[sorting=none,maxnames=6]{biblatex}
\usepackage{sectsty}
\allsectionsfont{ \normalfont} 

\usepackage{fancyhdr}
\numberwithin{equation}{section}		
\numberwithin{figure}{section}			
\numberwithin{table}{section}				

\newcommand{\horrule}[1]{\rule{\linewidth}{#1}} 	

\title{%
    \texorpdfstring{
		\usefont{OT1}{bch}{b}{n}
		\normalfont \normalsize \textsc{GSI Accelerator Physics Department} \\ [25pt]
		\horrule{0.5pt} \\[0.4cm]
		\huge Absence of Structure Resonances in SIS100 Tune Quadrant for Heavy Ion Fast Extraction \\
		\horrule{2pt} \\[0.5cm]
	}{%
	   Absence of Structure Resonances in SIS100 Tune Quadrant for Heavy Ion Fast Extraction%
	}%
}
\newcommand{\aujordhui}{20 October 2021}
\author{%
    \texorpdfstring{
		\normalfont 								\normalsize
        Adrian Oeftiger\footnote{a.oeftiger@gsi.de} \\[-3pt]
        \normalsize\textit{ GSI Helmholtzzentrum f\"ur Schwerionenforschung GmbH, Darmstadt, Germany} \\[-3pt]
        \normalsize
        \aujordhui \\[1em]
        \normalsize
        GSI Report: GSI-2021-01126
    }{%
        Adrian Oeftiger%
    }%
}
\date{\texorpdfstring{}{\aujordhui}}

\begin{document}
\maketitle

\begin{abstract}
    \emph{Abstract:} The understanding and avoidance of space charge induced resonances is of utmost importance for long storage times in synchrotrons, as they can lead to halo generation and subsequent beam loss. This report discusses the absence of structure resonances for heavy-ion operation in SIS100 in the tune quadrant foreseen for the fast extraction mode, $18.5<Q_{x,y}<19$ . Simulations of beam losses for the duration of the SIS100 accumulation plateau at nominal transverse space charge conditions with a maximum tune shift of $\Delta Q^{SC}_y=-0.3$ supplement the discussion.
\end{abstract}

\section{Setup of Beam Dynamics Simulations}
The full beam dynamics simulations of a single bunch in SIS100 with nonlinear space charge, nonlinear thin-lens tracking and nonlinear RF bucket are carried out using the open source CERN code SixTrackLib \cite{sixtracklib, sixtrack, sixtracklibgithub}.
The full SIS100 lattice \cite{syslatticerepo} is simulated based on the heavy-ion fast extraction tune settings, where the working point is scanned in $0.01$ intervals across the tunes $18.55\leq Q_{x,y}<19$.
The perturbation by the two normal-conducting radiation-hardened quadrupole magnets in the extraction insertion section will be mitigated by means of the SIS100 quadrupole corrector magnets to minimise the implied gradient error.
To simplify the setup, the lattice is assumed to perfectly satisfy the $S=6$ super-periodicity and all quadrupole magnets are taken to be of the cold, super-conducting type.

All beam and machine parameters are listed in Table \ref{tab: params}.

\begin{table}[htbp]
    \centering
    \caption{Parameters for Uranium-238 Beam Production in SIS100.}
    \renewcommand{\arraystretch}{1.3}
    \label{tab: params}
    \begin{tabular}{r|l}
        \textbf{Parameter} & \textbf{Value} \\ \hline
        Horizontal normalised rms emittance $\epsilon_x$ & \SI{5.9}{\milli\meter.mrad} \\
        Horizontal geometric KV emittance & \SI{35}{\milli\meter.mrad} \\
        Vertical normalised rms emittance $\epsilon_y$ & \SI{2.5}{\milli\meter.mrad} \\
        Vertical geometric KV emittance & \SI{15}{\milli\meter.mrad} \\
        Rms bunch length $\sigma_z$ & \SI{13.2}{\meter} \\
        Rms momentum deviation $\sigma_{\Delta p/p_0}$ & \SI{0.44d-3}{} \\
        Bunch intensity $N$ of U${}^{28+}_{238}$ & \SI{0.625d11}{} \\
        Max.\ space charge tune shift $\Delta Q^{SC}_y$ & $-0.30$ \\
        Chromatic tune spread $Q'_{x,y}\sigma_{\Delta p/p_0}$ & $0.02$ \\
        RF voltage (single-harmonic) $V_{RF}$ & \SI{58.2}{\kilo\volt} \\
        Harmonic $h$ & $10$ \\
        Kinetic energy & $E_{kin}=\SI{200}{\mega\electronvolt}/$u \\
        Relativistic $\beta$ factor & 0.568 \\
        Revolution frequency $f_{rev}$ & \SI{157}{\kilo\hertz} \\
    \end{tabular}
\end{table}

The simulations sample the bunch with 1000 macro-particles using an optics-matched 6D Gaussian phase space distribution.
This bunch is tracked through the synchrotron elements and space charge nodes, 501 of which are placed in intervals of slightly varying length along the ring.
Using more space charge nodes or more macro-particles does not significantly alter the simulation results.
Space charge is modelled as fixed (i.e.\ non-adaptive) frozen 3D Gaussian field maps which remain constant throughout the simulation, following the formula by Bassetti and Erskine \cite{bassetti,faddeeva}. 
The transverse beam size used in the field maps at each space charge node follows the computed local $\beta$- and dispersion functions and is based on the initial transverse emittances.
The transverse nonlinear space charge force is modulated with the longitudinal Gaussian beam profile. 
Evaluating the initial tune footprint in the simulation gives a maximum space charge tune shift from the bare working point in the vertical plane of $\Delta Q^{SC}_y=-0.3$ for the particles in the centre of the bunch.

\section{Simulations Results}

In a first baseline check, the storage of a zero-current beam in the symmetric lattice is simulated for the duration of the \SI{1}{\second} injection plateau.
The scanned tune quadrant exhibits zero beam loss and constant rms emittances -- in line with the absence of external resonance driving terms in the symmetric ideal lattice.

\begin{figure}[htb]
    \centering
    \includegraphics[width=0.6\linewidth]{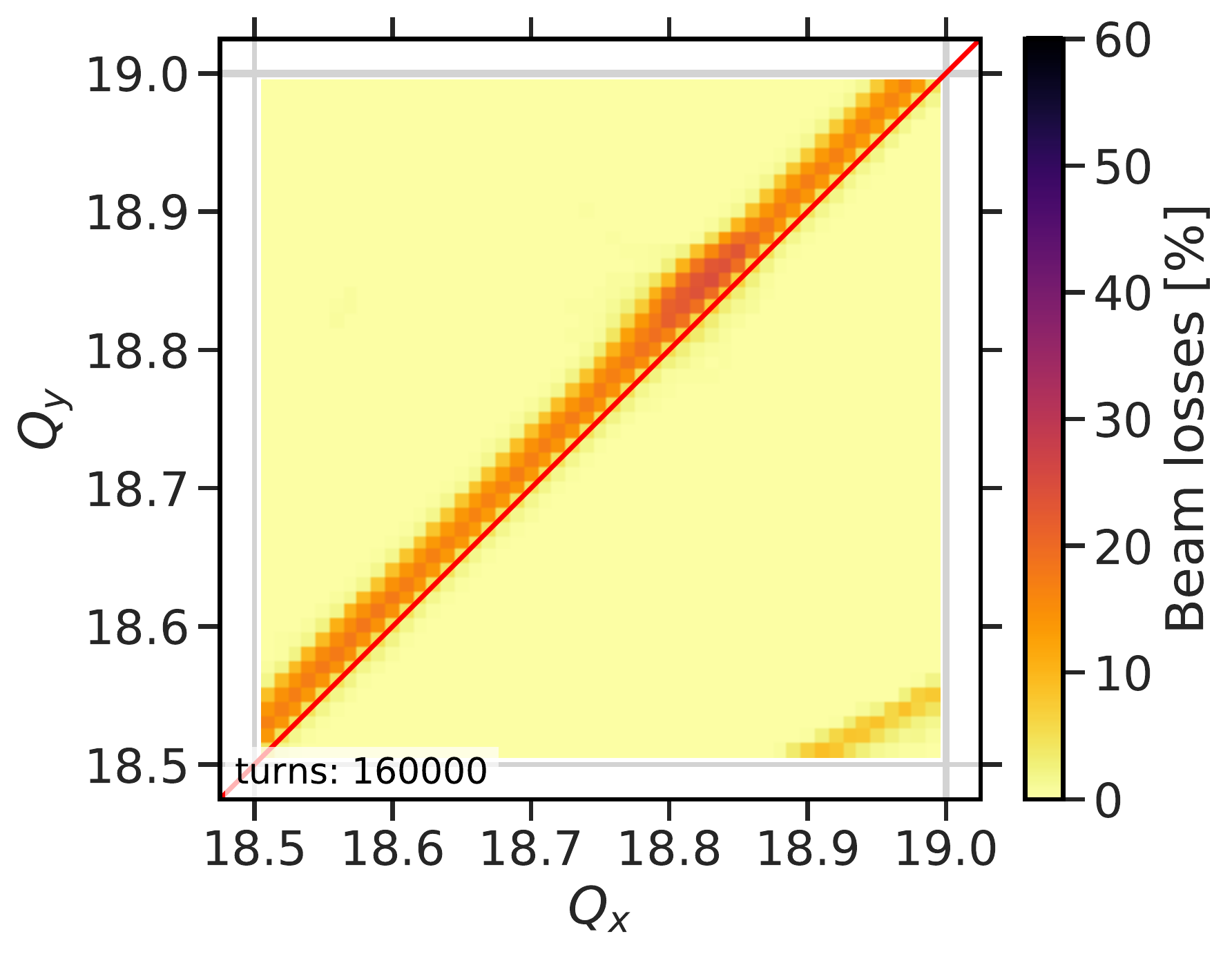}
    \caption{Symmetric cold lattice. Tune diagram with beam loss from FFSC simulations.}
    \label{fig: cold lattice}
\end{figure}

In a next step we include the fixed frozen space charge (``FFSC'') model in the simulations.
Simulating again the injection plateau duration in the symmetric lattice with FFSC for bunches of nominal intensity, the Montague stop-band \cite{montague1968fourth} appears around the coupling line $Q_x=Q_y$ as seen in Fig.~\ref{fig: cold lattice}.

It is important to note that no other resonances significantly limit the present tune quadrant based on the symmetric lattice.
The reason lies in the absence of low-order structure resonances crossing through the tune quadrant as opposed to e.g.\ the SIS18 ring \cite{hofmann2002space}.
This statement will be discussed in the following.

\section{Single-particle Resonance Condition}

The general (zero intensity) tune condition for betatron resonance \cite{courantsnyder} reads
\begin{equation} \label{eq: resonance condition}
    k\, Q_x + \ell\, Q_y = m
\end{equation}
for integer $k,\ell,m\in\mathbb{Z}$.
The order of the resonance is given by $n\doteq |k|+|\ell|$ while $m$ marks the harmonic driving the resonance.
Systematic structure resonances are resonances with a driving harmonic $m$ amounting to a multiple integer of structural symmetries in the ring lattice, such as the super-periodicity $S=6$ or the $84$ basic focusing cells.

\begin{figure}[htbp]
    \begin{subfigure}{0.48\linewidth}
        \centering
        \includegraphics[width=\linewidth]{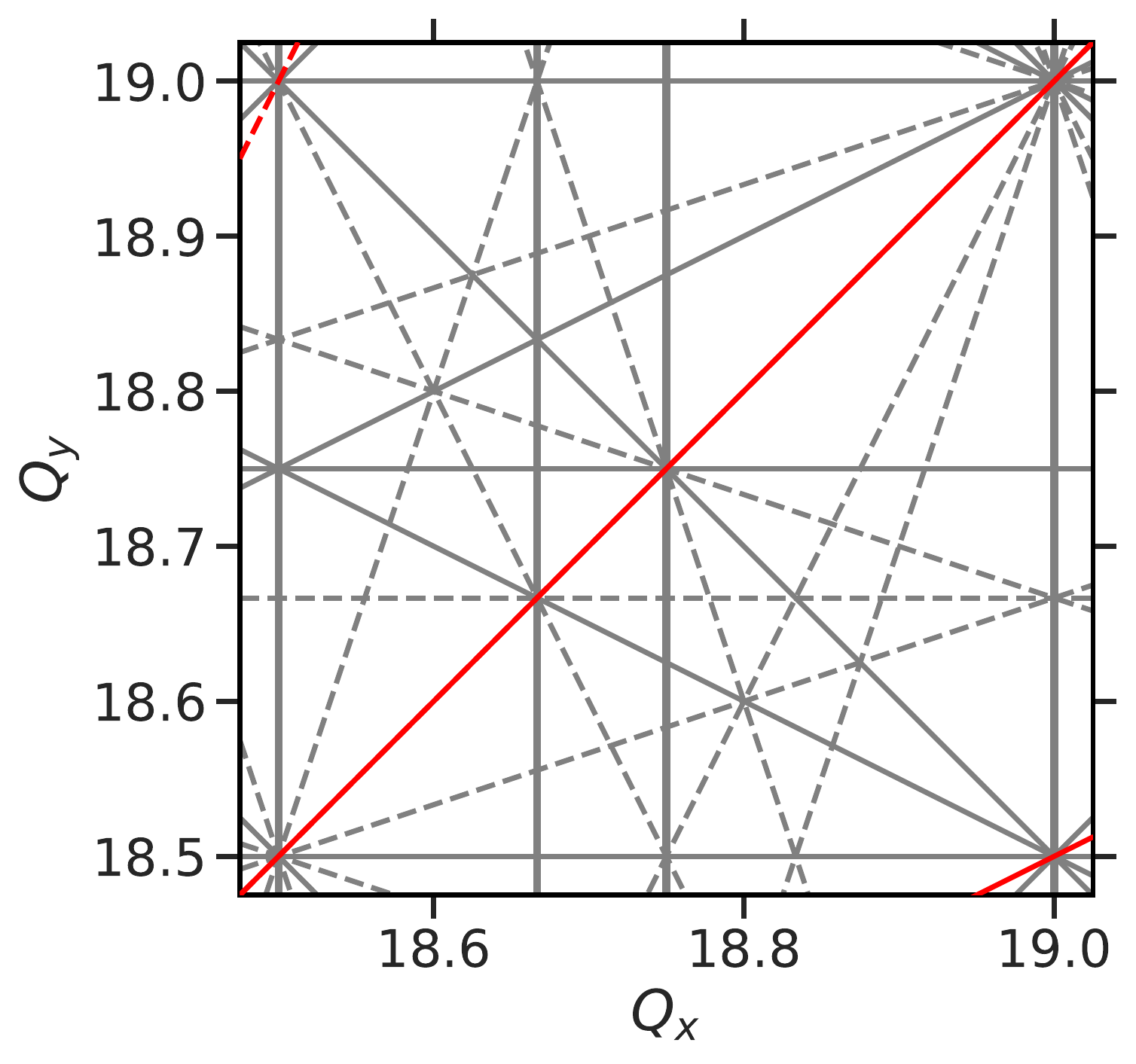}
        \caption{Up to octupolar order $n\leq 4$.}
        \label{fig: resonance diagrams oct}
    \end{subfigure}
    \hfill
    \begin{subfigure}{0.48\linewidth}
        \centering
        \includegraphics[width=\linewidth]{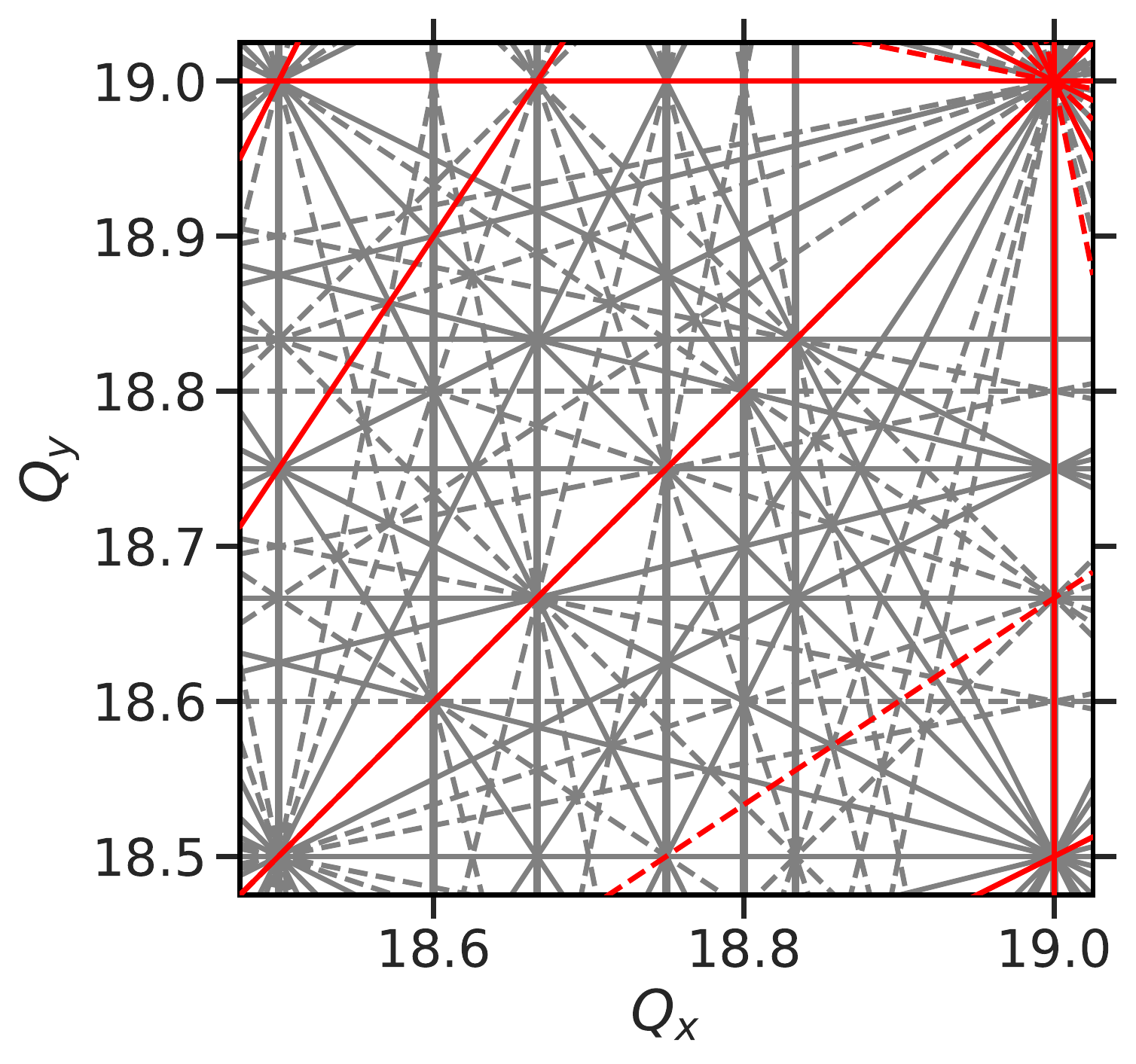}
        \caption{Up to dodecapolar order $n\leq 6$.}
        \label{fig: resonance diagrams dodeca}
    \end{subfigure}
    \\[2em] \centering
    \begin{subfigure}{\linewidth}
        \centering
        \includegraphics[width=0.48\linewidth]{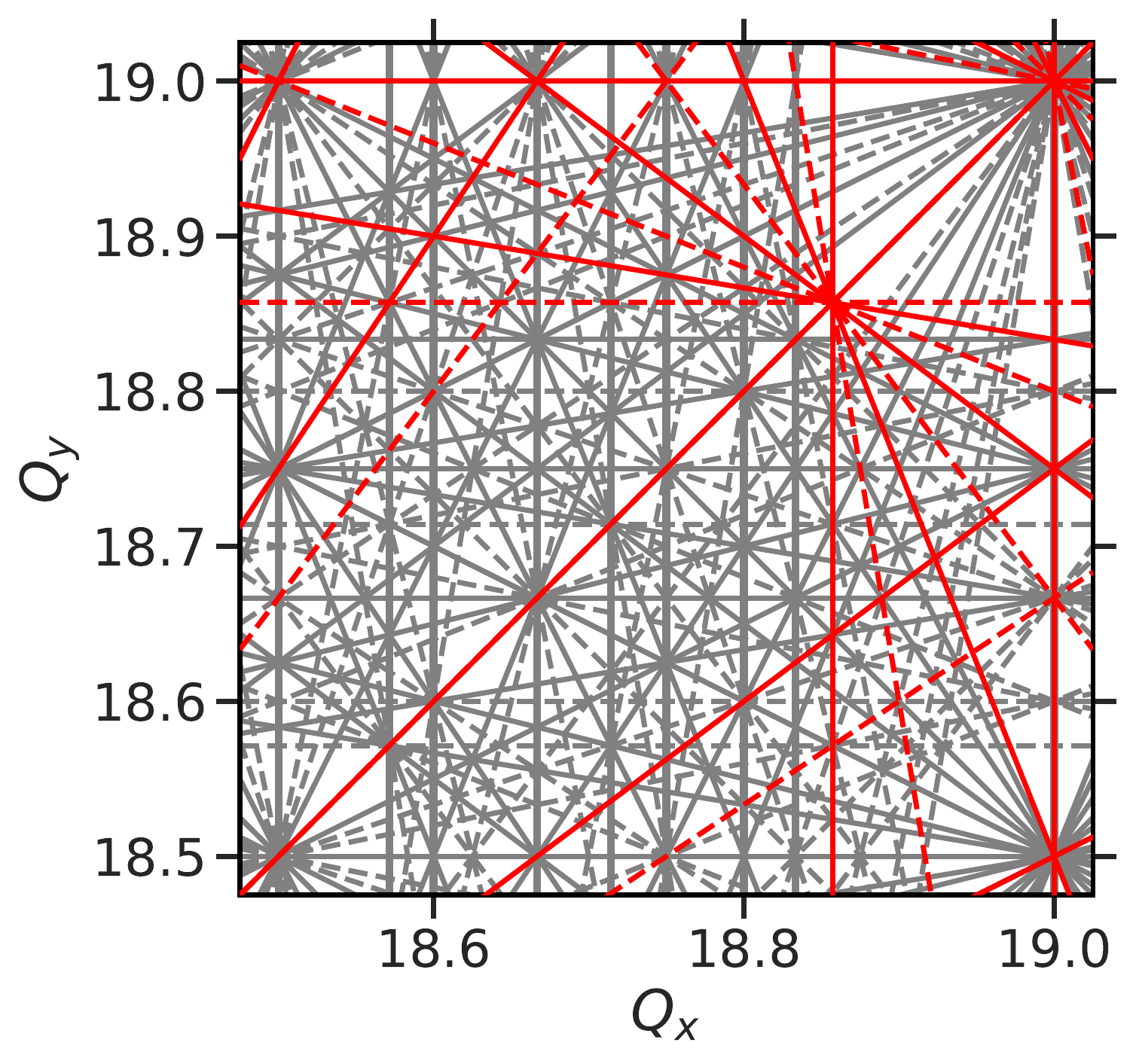}
        \caption{Up to order $n\leq 7$.}
    \end{subfigure}
    \caption{Resonance diagram for SIS100 with super-periodicity $S=6$ according to the single-particle Eq.~\eqref{eq: resonance condition}.}
    \label{fig: resonance diagrams}
\end{figure}

Figure \ref{fig: resonance diagrams} shows the lines where the zero-intensity resonance condition Eq.~\eqref{eq: resonance condition} is satisfied, where the red colour distinguishes the structure resonances with $m=r\,S$ for integer $r$.
Normal resonances driven by $b_n$ are plotted as solid lines, skew resonances driven by $a_n$ as dashed lines.
A normal and skew sextupole structure resonance appear on the periphery in the tune quadrant, going through $(Q_x,Q_y)=(19, 18.5)$ and $(18.5, 19)$, respectively.
For decapolar order $n=5$ we find one normal and one skew structure resonance above and below the coupling line, crossing further inside the tune diagram. 
Then, from order $n=7$ on and higher, the resonance diagram is crossed by many structure resonances.
Generally speaking it holds that the higher the order $n$, the weaker the resonance impact for equally strong $a_n,b_n$.

\section{Incoherent vs.\ Coherent Resonances}

The FFSC model can only predict incoherent resonances as the tracked particles cannot establish phase coherence.
For the considered SIS100 tune quadrant this is sufficient: coherent resonance phenomena appear to be irrelevant in the search for resonance-free areas, as the PIC simulations in \cite[Fig.~7a]{sis100beamloss} demonstrate.
In the following, we give a brief overview why this approach should suffice for typical situations in most space charge limited synchrotrons.

It is well known that space charge modifies the betatron resonance conditions Eq.~\eqref{eq: resonance condition} as recently discussed e.g.\ in \cite{PhysRevAccelBeams.24.024201} and references therein.
The particles experience a defocusing effect of the transverse space charge forces which imprints as an additional detuning term $\Delta Q_{x,y}^{SC}$ in this \emph{incoherent} resonance condition.
Furthermore, the space charge force leads to inter-particle communication and the particle distribution becomes sensitive to the excitation of coherent modes.
The coherent resonance condition of the corresponding bunch mode tunes involves intensity dependent $\mathcal{C}_n$ factors \cite{baartman1998betatron}, e.g.\ for a vertical $n$th order coherent resonance as
\begin{equation}\label{eq: coherent resonance condition}
    n\,\bigl(Q_y - \mathcal{C}_n\, \left|\Delta Q^{KV}_y\right|\bigr) = m \quad ,
\end{equation}
with the rms-equivalent Kapchinskij-Vladimirskij (KV) tune shift $\Delta Q^{KV}_y$ \cite{kv1959},
\begin{equation}
    Q^{KV}_y = -\frac{K^{SC} R^2}{4\sigma_y(\sigma_x + \sigma_y) Q_{y0}} \quad ,
\end{equation}
where $K^{SC}$ is the space charge perveance, $R$ the effective synchrotron radius, $\sigma_{x,y}$ the average rms beam sizes in the transverse plane and $Q_{y0}$ the bare tune.

The structure resonance condition of integral $m=r\,S$ (where $r\in \mathbb{Z}$) for external driving terms translates to internally driven parametric resonance in the presence of space charge.
They can have a severe impact on beam quality when a pumped instability occurs.
For the pure alternate gradient focusing (without space charge) this parametric resonance mechanism is known from the single-particle $180^\circ$ or Mathieu instability.
For finite space charge also lower phase advance than $180^\circ$ per cell can lead to parametric resonance, specifically when coherent beam modes become unstable.
An important feature of the parametric resonance condition is that this instability occurs twice as dense in tune space compared to the regular resonances in Eqs.~\eqref{eq: resonance condition} and \eqref{eq: coherent resonance condition}, namely at half-integer harmonics $m\mapsto \frac{m}{2}$.
While in parametric resonance research this insight goes as far back as to the 19th century (cf.\ e.g.\ Ref.~\cite{rayleigh1887xvii}), the occurrence of the $m/2$ condition in intense beam dynamics is known at least since Ref.~\cite{hofmann1982stability}.

Recently, Ref.~\cite{PhysRevAccelBeams.22.074201} argued that the 2D resonance diagram should be constructed entirely based on parametric coherent resonances including the nonlinear orders, which would mean twice as many red lines in Fig.~\ref{fig: resonance diagrams}, based on the half-integer harmonic condition $\frac{m}{2}=r\,S$.
Fortunately, for Gaussian bunch distributions in alternate gradient focusing, parametric coherent resonance occurs only for the second-order $n=2$ case (also called the envelope instability).
As noted for 1D in Ref.~\cite{Okamoto:2002zm} and for 2D (and bunched beam) in Ref.~\cite{PhysRevAccelBeams.24.024201}, nonlinear orders $n\geq 3$ are Landau damped.
Therefore, higher-order parametric coherent resonances seem to be generally absent for realistic beam conditions.

For SIS100 with $S=6$, the nearest $n=2$ parametric coherent resonance stop-bands are located close to (above) $Q_{x,y}=18$ and $Q_{x,y}=19.5$, and thus far away from the design tune quadrant $18.5\leq Q_{x,y} \leq 19$ for fast extraction of heavy-ion beams.
Even when searching for loss-free working points close to such an envelope instability stop-band, the results of Ref.~\cite{PhysRevAccelBeams.24.024201} show that the incoherent 4th order stop-band overlaps and entirely embraces the coherent one -- both on the lower end (halo tune region) and the higher end (outer core tune region). 
The bunched beam FFSC predictions are found to correctly identify the resonance-free tune space.

Also in the absence of parametric coherent resonance (e.g.\ due to Landau damping), structure resonances at integer harmonics $m=r\,S$ can be harmful: the incoherent resonance mechanism can generally lead to halo formation and eventual beam loss.
Driving terms can be provided externally (e.g.\ regularly spaced sextupole magnets) or internally (e.g.\ space charge).
In the latter case, the modulation of the space charge potential along the alternate gradient focusing provides resonance driving terms for all even-order $k,\ell=0,2,4,...$ in Eq.~\eqref{eq: resonance condition}.
Large-amplitude particles meeting this incoherent resonance condition are excited and potentially driven into the machine aperture.
The CERN Proton Synchrotron is an example for space charge limitation by the halo generation through an 8th order structure resonance \cite{PhysRevAccelBeams.23.091001}, demonstrating that even very high orders (which are usually disregarded) can be detrimental.

In summary, these incoherent resonance mechanisms for halo and outer core are caught by the FFSC model when using realistic lattices. 
It is very useful to understand that computational prediction of the location of incoherent space charge driven stop-bands with FFSC are hence not only valid, but even a viable means to identify the edges of resonance-free areas.

\section{Identifying the Incoherent Space Charge Driven Resonances in SIS100}
We turn our attention back to the SIS100 FFSC simulations presented in Fig.~\ref{fig: cold lattice}. 
Driving terms for nonlinear resonances are solely provided by the space charge potential, where the harmonic $m$ is determined by the beam size modulation given by the structure of the lattice. 
To stress it once more, only structure resonances with even $k,\ell$ can therefore appear in the tune diagram.
The predicted beam loss appearing in Fig.~\ref{fig: resonance diagrams oct} close to the third-order structure resonance $Q_x-2Q_y=-18$, which touches the tune quadrant at $Q_x=19$ and $Q_y=18.5$, is in reality induced by the coinciding sixth-order structure resonance \begin{equation}
    2Q_x-4Q_y=-36=-6\cdot S \quad .
\end{equation}
The next higher-order structure resonances in Fig.~\ref{fig: resonance diagrams dodeca} are the decapolar ones: the normal one above the coupling line, satisfying $3Q_x-2Q_y=18$, shows nearby weak rms emittance exchange as well as minimal loss on the $1\%$ level in the FFSC simulations.
Again, the actual reason for the observed dynamics is the space charge driven 10th order structure resonance 
\begin{equation}
    6Q_x-4Q_y=36=6\cdot S \quad .
\end{equation} 
Its effect of $\approx 1\%$ beam loss is already too weak to appear on the beam loss scale shown in Fig.~\ref{fig: cold lattice}. 

\begin{figure}[htb]
    \centering
    \includegraphics[width=0.6\linewidth]{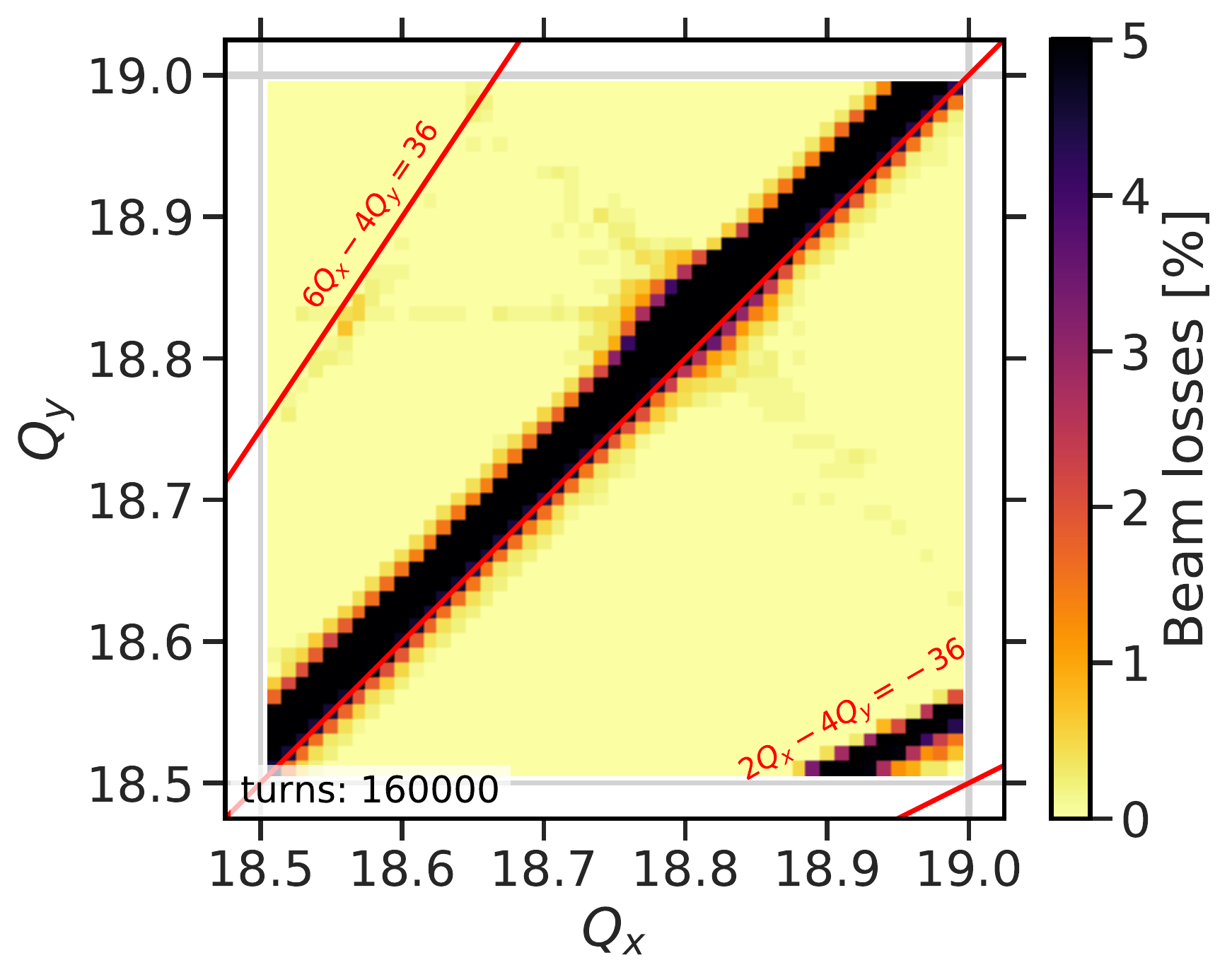}
    \caption{Same as Fig.~\ref{fig: cold lattice} with structure resonances indicated.}
    \label{fig: cold lattice details}
\end{figure}

Figure \ref{fig: cold lattice details} shows the same beam loss results with a smaller scale of up to $5\%$ beam loss, with the two observed space charge driven structure resonances indicated in red.
Further very low losses are visible here along the 14th order resonance lines.
However, all these high-order resonances with $1\%$ beam loss and less play no role in realistic scenarios including the magnet imperfection driven resonances -- compare e.g.\ to 
FFSC simulations with the SIS100 field error model as presented in \cite[Fig.~7a]{sis100beamloss}, where the usual externally driven resonances simply overshadow the 10th order space space charge driven resonances given the SIS100 design beam parameters.

\section{Conclusion}

All in all, from the optics design point of view, one should avoid tune quadrants with structure resonances to push the space charge limit in synchrotrons with an accumulation plateau.
At the same time, it is not necessary to consider half-integer harmonics for structure resonances due to the parametric coherent resonances -- one can resort to the usual incoherent resonance diagrams to at least 6th order.
The incoherent stopbands will embrace and include the coherent stopbands as long as the space charge induced tune spread remains unmodified (which can happen e.g.\ due to nonlinear electron lenses).
The lesson learned is that it is sufficient to search for incoherent resonance phenomena with fixed frozen space charge models in order to correctly identify resonance-free areas.

\printbibliography

\end{document}